\documentclass[zpreprint,zbstnp]{zeus_paper}

\usepackage[english]{babel}

\chardef\usc=95
\chardef\til=126
\catcode`\@=11 
\DeclareRobustCommand\xdotspace{\futurelet\@let@token\@xdotspace}
\def\@xdotspace{%
  \ifx\@let@token.\else
  \ifx\@let@token\bgroup.\else
  \ifx\@let@token\egroup.\else
  \ifx\@let@token\/.\else
  \ifx\@let@token\ .\else
  \ifx\@let@token~.\else
  \ifx\@let@token!.\else
  \ifx\@let@token,.\else
  \ifx\@let@token:.\else
  \ifx\@let@token;.\else
  \ifx\@let@token?.\else
  \ifx\@let@token/.\else
  \ifx\@let@token'.\else
  \ifx\@let@token).\else
  \ifx\@let@token-.\else
  \ifx\@let@token\@xobeysp.\else
  \ifx\@let@token\space.\else
  \ifx\@let@token\@sptoken.\else
   .\space
   \fi\fi\fi\fi\fi\fi\fi\fi\fi\fi\fi\fi\fi\fi\fi\fi\fi\fi}
\catcode`\@=12 

\newcommand{\stru}[2]{%
   \relax\ifmmode\hbox{\vrule height#1 depth#2 width0pt}%
   \else\vrule height#1 depth#2 width0pt\fi}

\newcommand{\Ronum}[1]{\uppercase\expandafter{\romannumeral#1}}
\newcommand{\ronum}[1]{\expandafter{\romannumeral#1}}
\DeclareRobustCommand{\LaTeXZ}{%
  \LaTeX\kern-.05em4\kern-.1em
  {\raisebox{-0.2ex}{$\scriptstyle\text{ZEUS}$}}\xspace}

\DeclareMathAlphabet{\mathbf}{OT1}{cmr}{bx}{sl}
\newcommand{\eVdist}{\kern-0.06667em}

\newcommand{\gev}{{\,\text{Ge}\eVdist\text{V\/}}}

\newcommand{\Tesla}{\,\text{T}}

\newcommand{\slashfrac}[2]{%
  \raisebox{0.5ex}{\ensuremath #1}\kern-0.12em/\kern-0.08em
  \raisebox{-.8ex}{\ensuremath #2}}

\newcommand{\sqr}[3]{%
    {\vcenter{\hrule height.#3ex\hbox{\vrule width.#2ex height#1ex
     \kern#1ex\vrule width.#3ex}\hrule height.#2ex}}}

\catcode`\@=11 
\newcommand{\parenbar}{\mathpalette\p@renb@r}
\def\p@renb@r#1#2{\vbox{%
  \ifx#1\scriptscriptstyle \dimen@.7em\dimen@ii.2em\else
  \ifx#1\scriptstyle \dimen@.8em\dimen@ii.25em\else
  \dimen@1em\dimen@ii.4em\fi\fi \offinterlineskip
  \ialign{\hfill##\hfill\cr
    \vbox{\hrule width\dimen@ii}\cr
    \noalign{\vskip-.3ex}%
    \hbox to\dimen@{$\mathchar300\hfil\mathchar301$}\cr
    \noalign{\vskip-.3ex}%
    $#1#2$\cr}}}
\catcode`\@=12 

\newcommand{\IP}{{\rm I$\kern-0.01667em$P}\xspace}

\mathchardef\qsm=63
\mathchardef\pls=43
\mathchardef\mns=512
\mathchardef\plm=518
\mathchardef\eql=61
\mathchardef\smallleft=300
\mathchardef\smallright=301
\mathchardef\les=316
\mathchardef\gre=318
\mathchardef\leq=532
\mathchardef\grq=533

\catcode`\@=11 
\newcounter{pict@width}
\newcounter{pict@height}
\newlength{\pict@scale}
\newcommand{\psfigadd}[4]{%
\setcounter{pict@width}{1*\ratio{#2+\pict@scale/2}{\pict@scale}}
\setcounter{pict@height}{1*\ratio{#3+\pict@scale/2}{\pict@scale}}
\setlength{\unitlength}{\pict@scale}
\hbox to #2{\hspace{-\fill}\begin{picture}(\thepict@width,\thepict@height)
\put(0,0){\psfig{figure=#1,width=#2,height=#3,clip=}}
\SetScale{0.283466457}
\SetWidth{1.763889}
{#4}
\end{picture}}
}
\newcounter{pict@widthfst}
\newcounter{pict@widthscd}
\newcounter{pict@widthtot}
\newcommand{\psfigaddtwo}[7]{%
\setcounter{pict@widthfst}{1*\ratio{#2+\pict@scale/2}{\pict@scale}}
\setcounter{pict@widthscd}{1*\ratio{#2+#4+\pict@scale/2}{\pict@scale}}
\setcounter{pict@widthtot}{1*\ratio{#2+#4+#6+\pict@scale/2}{\pict@scale}}
\setcounter{pict@height}{1*\ratio{#3+\pict@scale/2}{\pict@scale}}
\setlength{\unitlength}{\pict@scale}
\hbox{\hspace{-\fill}\begin{picture}(\thepict@widthtot,\thepict@height)
\put(0,0){\psfig{figure=#1,width=#2,height=#3,clip=}}
\put(\thepict@widthscd,0){\psfig{figure=#5,width=#6,height=#3,clip=}}
\SetScale{0.283466457}
\SetWidth{1.763889}
{#7}
\end{picture}}
}
\newcommand{\psfigror}[4]{%
\setcounter{pict@width}{1*\ratio{#2+\pict@scale/2}{\pict@scale}}
\setcounter{pict@height}{1*\ratio{#3+\pict@scale/2}{\pict@scale}}
\setlength{\unitlength}{\pict@scale}
\hbox{\begin{picture}(\thepict@width,\thepict@height)
\put(0,\thepict@height){\psfig{figure=#1,width=#3,height=#2,clip=,angle=270}}
\SetScale{0.283466457}
\SetWidth{1.763889}
{#4}
\end{picture}}
}
\newcommand{\psfigrol}[4]{%
\setcounter{pict@width}{1*\ratio{#2+\pict@scale/2}{\pict@scale}}
\setcounter{pict@height}{1*\ratio{#3+\pict@scale/2}{\pict@scale}}
\setlength{\unitlength}{\pict@scale}
\hbox{\begin{picture}(\thepict@width,\thepict@height)
\put(0,0){\psfig{figure=#1,width=#3,height=#2,clip=,angle=90}}
\SetScale{0.283466457}
\SetWidth{1.763889}
{#4}
\end{picture}}
}
\catcode`\@=12 
\newlength\listtextwidth

\catcode`\@=11 
\newlength{\@tabfninsert}
\newlength{\@tabfnwidth}
\newcommand{\tabfootnote}[2]{%
  \setlength{\@tabfninsert}{0.8em}
  \setlength{\@tabfnwidth}{\textwidth}
  \addtolength{\@tabfnwidth}{-\@tabfninsert}
  \addtolength{\@tabfnwidth}{-0.4em}
  \noindent\makebox[\@tabfninsert][r]{\footnotesize$^{#1}$\hfil}\hfill%
  \parbox[t]{\@tabfnwidth}{\footnotesize #2\hfill}}
\catcode`\@=12 

\def\citeCTD{{\cite{%
nim:a279:290,*npps:b32:181,*nim:a338:254%
}}\xspace}

\def\citeCAL{{\cite{%
nim:a309:77,*nim:a309:101,*nim:a321:356,*nim:a336:23%
}}\xspace}

\begin{document}

\prepnum{DESY--08--089\\ZEUS-pub-08-005}

\title{
Search for events with an isolated lepton and missing transverse  momentum and a measurement of $\mathbf{W}$ production at HERA
}                                                       
                    
\author{ZEUS Collaboration}
\date{July 2008}

\abstract{
A search for events with an isolated high-energy lepton and large 
missing transverse momentum has been performed with the ZEUS detector 
at HERA using a total integrated luminosity of 504\,$\mathrm{pb}^{-1}$. The 
results agree well with  Standard Model predictions. The cross section for 
 production of single $W$ bosons in electron-proton collisions with unpolarised
 electrons is measured to be 
$0.89^{+0.25}_{-0.22}$\,(stat.)\,$\pm{0.10}$\,(syst.)\,pb.
}

\makezeustitle

\pagenumbering{Roman}

%
%
%
%
                                                   %
\begin{center}                                                                                     
{                      \Large  The ZEUS Collaboration              }                               
\end{center}                                                                                       
  S.~Chekanov,                                                                                     
  M.~Derrick,                                                                                      
  S.~Magill,                                                                                       
  B.~Musgrave,                                                                                     
  D.~Nicholass$^{   1}$,                                                                           
  \mbox{J.~Repond},                                                                                
  R.~Yoshida\\                                                                                     
 {\it Argonne National Laboratory, Argonne, Illinois 60439-4815, USA}~$^{n}$                       
\par \filbreak                                                                                     
  M.C.K.~Mattingly \\                                                                              
 {\it Andrews University, Berrien Springs, Michigan 49104-0380, USA}                               
\par \filbreak                                                                                     
  P.~Antonioli,                                                                                    
  G.~Bari,                                                                                         
  L.~Bellagamba,                                                                                   
  D.~Boscherini,                                                                                   
  A.~Bruni,                                                                                        
  G.~Bruni,                                                                                        
  F.~Cindolo,                                                                                      
  M.~Corradi,                                                                                      
\mbox{G.~Iacobucci},                                                                               
  A.~Margotti,                                                                                     
  R.~Nania,                                                                                        
  A.~Polini\\                                                                                      
  {\it INFN Bologna, Bologna, Italy}~$^{e}$
% ...
\newpage                                                                                           
\enlargethispage{5cm}                                                                              
$^{\    1}$ also affiliated with University College London,                                        
United Kingdom\\                                                                                   
% ...
$^{\    5}$ now at University of Bonn, Germany \\                                                  
% ...
$^{  10}$ now at BayesForecast, Madrid, Spain \\                                                   
% ...
%
\newpage   
                                                           %
                                                           %
\begin{tabular}[h]{rp{14cm}}                                                                       
$^{a}$ &  supported by the Natural Sciences and Engineering Research Council of Canada (NSERC) \\  
$^{b}$ &  supported by the German Federal Ministry for Education and Research (BMBF), under        
          contract numbers 05 HZ6PDA, 05 HZ6GUA, 05 HZ6VFA and 05 HZ4KHA\\                         
$^{c}$ &  supported in part by the MINERVA Gesellschaft f\"ur Forschung GmbH, the Israel Science   
          Foundation (grant no. 293/02-11.2) and the U.S.-Israel Binational Science Foundation \\  
$^{d}$ &  supported by the Israel Science Foundation\\                                             
$^{e}$ &  supported by the Italian National Institute for Nuclear Physics (INFN) \\                
$^{f}$ &  supported by the Japanese Ministry of Education, Culture, Sports, Science and Technology 
          (MEXT) and its grants for Scientific Research\\                                          
$^{g}$ &  supported by the Korean Ministry of Education and Korea Science and Engineering          
          Foundation\\                                                                             
$^{h}$ &  supported by the Netherlands Foundation for Research on Matter (FOM)\\                   
$^{i}$ &  supported by the Polish State Committee for Scientific Research, project no.             
          DESY/256/2006 - 154/DES/2006/03\\                                                        
$^{j}$ &  partially supported by the German Federal Ministry for Education and Research (BMBF)\\   
$^{k}$ &  supported by RF Presidential grant N 8122.2006.2 for the leading                         
          scientific schools and by the Russian Ministry of Education and Science through its      
          grant for Scientific Research on High Energy Physics\\                                   
$^{l}$ &  supported by the Spanish Ministry of Education and Science through funds provided by     
          CICYT\\                                                                                  
$^{m}$ &  supported by the Science and Technology Facilities Council, UK\\                         
$^{n}$ &  supported by the US Department of Energy\\                                               
$^{o}$ &  supported by the US National Science Foundation. Any opinion,                            
findings and conclusions or recommendations expressed in this material                             
are those of the authors and do not necessarily reflect the views of the                           
National Science Foundation.\\                                                                     
$^{p}$ &  supported by the Polish Ministry of Science and Higher Education                         
as a scientific project (2006-2008)\\                                                              
$^{q}$ &  supported by FNRS and its associated funds (IISN and FRIA) and by an Inter-University    
          Attraction Poles Programme subsidised by the Belgian Federal Science Policy Office\\     
$^{r}$ &  supported by the Malaysian Ministry of Science, Technology and                           
Innovation/Akademi Sains Malaysia grant SAGA 66-02-03-0048\\                                       
\end{tabular}                                                                                      
                                                           %
\newpage                                                           %

\pagenumbering{arabic} 
\pagestyle{plain}

\section{Introduction}

The production of $W$ bosons in electron\footnote{In this paper
  ``electron'' refers both to electrons and positrons unless stated
  otherwise.}-proton ($ep$) collisions is an interesting Standard
  Model (SM) process with a small cross section. This process, with
  subsequent leptonic decay of the $W$ boson, also constitutes one of
  the most important SM backgrounds to many new physics
  searches~\cite{pl:b559:153,pl:b561:241}, for which high-energy
  leptons and large missing transverse momentum,
  $P_T^{\mathrm{miss}}$, are common signatures. Such searches have
  been performed previously by both the
  H1\cite{epj:c5:575,pl:b561:241,epj:c48:699} and ZEUS
  \cite{pl:b471:411,pl:b559:153,pl:b583:41} collaborations. The H1
  collaboration observed an excess of electron or muon events with
  large hadronic transverse momentum, $P_T^X$, over the SM
  predictions.  Previous ZEUS results have not confirmed this excess.

This paper presents a new search and a measurement of the cross
section for $W$ production at HERA.  The study was performed by
selecting events containing isolated electrons or muons with high
transverse momentum, $P_T^l$, in events with large
$P^{\mathrm{miss}}_T$. The data used were taken from 1994 to 2007. 
 The total integrated luminosity analysed was
$504\,\mathrm{pb^{-1}}$, a four-fold increase compared to previous
ZEUS searches \cite{pl:b559:153,pl:b583:41}.

\section{Standard Model expectations}
\label{sec:SMexp}
The SM predicts the production of single $W$ and $Z$ bosons in $ep$
collisions at HERA. The event topology studied here is large
$P_T^{\mathrm{miss}}$ and an isolated lepton with large $P_T^l$.

\begin{description}
\item \ \ \ {$\mathbf{W}$\bf  production}: $ep \rightarrow eWX$ or $ep \rightarrow \nu WX$\\ 
Neutral current $W$ production, $ep \rightarrow eWX$, with subsequent
 leptonic decay, $W{\rightarrow}l\nu$, is the dominant SM process that
 produces events matching the desired topology.  Charged current $W$
 production, $ep \rightarrow \nu WX$, with subsequent leptonic decay
 also produces such events.

 The SM production cross section, obtained from a calculation
      including Quantum Chromodynamics (QCD) corrections at
      next-to-leading-order (NLO)~\cite{epj:c25:405,jp:g25:1434}, is
      1.1\,pb and 1.3\,pb for the relevant centre-of-mass energies,
      $\sqrt{s}$, of 300\,GeV and 318\,GeV respectively. The estimated
      uncertainty on this calculation is 15\%.  Monte Carlo (MC)
      events have been generated with the leading-order {\sc Epvec}
      generator \cite{np:b375:3} and weighted by a factor dependent on
      the transverse momentum and rapidity of the $W$, such that the
      resulting cross sections correspond to the NLO calculation
      \cite{hep-ex/0302040}.  The {\sc Epvec} MC is also used to
      generate $ep \rightarrow \nu_eWX$ events.  The contribution of
      $ep \rightarrow \nu_eWX$ to the total $W$ production cross
      section is approximately $7\%$.

\item \ \ \ {$\mathbf{Z}$ \bf  production}: $ep \rightarrow eZ(\rightarrow \nu\bar{\nu})X$\\ 
The process $ep \rightarrow eZ(\rightarrow \nu\bar{\nu})X$ can produce
high-energy scattered electrons and large $P_T^{\mathrm{miss}}$. The
visible cross section for this process as calculated by {\sc{Epvec}}
is less than 3\% of the predicted $W$ production cross section. It was
neglected in this analysis.

\end{description}

 Several SM processes can produce events with large $P_T^{\mathrm{miss}}$ and
high-energy leptons as a result of mismeasurements.

\begin{description}
\item \ \ \ {\bf Neutral current deep inelastic scattering (NC DIS)}:  $ep \rightarrow eX$\\
Genuine isolated high-energy electrons are produced in NC DIS.
Together with a fake $P_T^{\mathrm{miss}}$ signal from mismeasurement,
they form the dominant fake signal in searches for isolated electrons
at high $P_T^X$. Neutral current DIS events were simulated using the
generator {\sc Django6} \cite{cpc:81:381}, an interface to the MC
programs {\sc Heracles} 4.5 \cite{cpc:69:155} and {\sc Lepto} 6.5
\cite{cpc:101:108}. Leading-order electroweak radiative corrections
were included and higher-order QCD effects were simulated using the
colour-dipole model of {\sc Ariadne~4.08}
\cite{cpc:71:15}. Hadronisation of the partonic final state was
performed by {\sc Jetset} \cite{cpc:39:347}.

\item \ \ \ {\bf Charged current deep inelastic scattering (CC DIS)}:  $ep \rightarrow \nu X$\\
A CC DIS event can mimic the selected topology if it contains a fake
electron as there is real $P_T^{\mathrm{miss}}$ due to the escaping
neutrino.  Charged current DIS events were simulated using the
generator {\sc Django6} as described for the NC DIS events.

\item \ \ \ {\bf Lepton pair production}: $ep \rightarrow e l^+l^-X$\\
Lepton pair production can mimic the selected topology if one lepton
escapes detection or measurement errors cause apparent missing
momentum.  Lepton pair production is the dominant fake signal in
searches for isolated high-$P_T$ muons. This process was simulated
using the {\sc Grape} \cite{cpc:136:126} dilepton generator.

\item \ \ \ {\bf Photoproduction of jets}:  $\gamma p \rightarrow X$\\
Hard photoproduction processes can also contribute to the fake signal
rate. This may occur when a particle from the hadronic final state is
interpreted as an isolated lepton together with a fake
$P^{\mathrm{miss}}_T$ signal arising from
mismeasurement. Photoproduction processes as simulated with {\sc
Herwig}~6.1~\cite{cpc:67:465} make a negligible contribution to the SM
expectation.
\end{description}

\section{The ZEUS detector}
A detailed description of the ZEUS detector can be found
elsewhere~\cite{zeus:1993:bluebook}. Charged particles were tracked in
the central tracking detector (CTD)~\citeCTD which operated in a
magnetic field of $1.43\,\Tesla$ provided by a thin superconducting
solenoid. Before the 2003--2007 running period, the ZEUS tracking
system was upgraded with a silicon micro vertex detector
(MVD)~\cite{nim:a581:656}.  The high-resolution uranium--scintillator
calorimeter (CAL)~\citeCAL consisted of three parts: the forward, the
barrel and the rear calorimeters. The smallest subdivision of the CAL
was called a cell.  A three-level trigger was used to select events
online \cite{uproc:chep:1992:222,*nim:a580:1257} requiring large
$P_T^{\mathrm{miss}}$ or well isolated electromagnetic deposits in the
CAL.

\section{Event reconstruction}

\label{sec:presel}

Electrons were identified by an algorithm that selects candidate
 electromagnetic clusters in the CAL and combines them with tracking
 information. The algorithm was optimised for maximum electron-finding
 efficiency and electron-hadron separation for NC DIS
 events~\cite{zfp:c74:207}.  Electromagnetic clusters were classified
 as isolated electron candidates when the energy not associated with
 the cluster in an $\{\eta,\phi\}$ cone of radius 0.8 around the
 electron direction was less than $5\gev$ and less than $5\%$ of the
 electromagnetic cluster energy measured with the calorimeter, where
 $\eta = - \log ( \tan (\theta /2))$.

Muons were identified through their signature as minimum ionising
particles (MIPs). Their energy depositions can be spread over several
calorimeter clusters.  Therefore, neighbouring clusters were grouped
together into larger-scale objects which, provided they passed
topological and energy cuts, were classified as CAL MIPs. In this
analysis a muon candidate was selected if a CAL MIP matched an
extrapolated CTD track from the primary vertex to within 20 cm.

The missing transverse momentum was determined from calorimetric and
tracking information.  The magnitude of the missing transverse
momentum measured with the CAL was defined as

\begin{equation}
P^{\mathrm{CAL}}_T = \sqrt{\left( \sum \limits_i p^{\mathrm{CAL}}_{X,i} \right)^2+ \left( \sum \limits_i p^{\mathrm{CAL}}_{Y,i} \right)^2}, \nonumber
\end{equation}

where $p^{\mathrm{CAL}}_{X,i}=E_i \sin{\theta_i} \cos{ \phi_i}$ and
$p^{\mathrm{CAL}}_{Y,i}= E_i \sin{\theta_i}\sin{\phi_i}$ were
calculated from individual energy deposits, $E_i$, in clusters of CAL
cells corrected\cite{epj:c11:427} for energy loss in inactive
material.  In $W \rightarrow e \nu$ events, $P^{\mathrm{CAL}}_T$ as
defined above is an estimate of the missing transverse momentum
carried by the neutrino, $P_T^{\nu}$. In $W
\rightarrow \mu \nu$ events, the muon
deposits very little energy in the calorimeter and therefore a better
estimate of $P_T^{\nu}$ can be
obtained if the momentum of the muon is calculated from its track
measured in the CTD ($p^{\mu,\mathrm{track}}$). Combination with the above
estimate of the total transverse momentum from the calorimeter leads to
\begin{equation}
P_T^{\rm{miss}} = \sqrt{\left( \sum \limits_i p_{X,i}^{\rm{CAL}}+p_{X}^{\mu,\rm{track}} \right)^2+ \left( \sum \limits_i p_{Y,i}^{\rm{CAL}}+p_{Y}^{\mu,\rm{track}} \right)^2}. \nonumber
\end{equation}
The hadronic transverse momentum, $P_T^X$, was defined as the sum over those
calorimeter cells that are not assigned to lepton-candidate
clusters.

 The charged-lepton transverse momentum, $P_T^l$, was calculated from
the calorimeter cluster for $l=e$ and from the track momentum for
$l=\mu$.  The transverse mass for $W$ bosons decaying via $W
\rightarrow l \nu$ is defined as:
\begin{equation}
M_T=\sqrt{2 P^{l}_{T} P^{\nu}_T (1-\cos{\phi^{l\nu}})}, \nonumber
\end{equation}
where  $\phi^{l \nu}$ is the azimuthal separation of the lepton and 
$P_T^{\nu}$ vectors.

The following event properties were used to suppress backgrounds from
mismeasured large $P_T^{\mathrm{miss}}$ and fake high-energy
leptons. Selection cuts on these event properties will be briefly
described in Section \ref{sec:elec}.

The quantity $\xi^2_e$ was defined as
\begin{equation}
 \xi_e^2 = 2E'_eE_e(1+\cos \theta_e), \nonumber
\end{equation}
where $E'_e $ is the energy of the final-state electron, $E_{e} =
27.5$\,GeV is the electron beam energy and $\theta_e$ is the polar
angle of the electron measured in the calorimeter. For NC DIS events,
where the scattered electron is identified as the isolated lepton,
$\xi_e^2$ corresponds to the virtuality of the exchanged boson,
$Q^2$. Neutral current DIS events generally have low values of
$\xi^2_e$ whilst electrons from $W$ decay will generally have high
values of $\xi^2_e$.

The acoplanarity angle, $\phi_{\mathrm{acop}}$, is the azimuthal
separation in the $\{X,Y \}$ plane of the outgoing lepton and the
vector that balances the hadronic transverse momentum vector. For well
measured NC DIS events, $\phi_{\mathrm{acop}}$ is close to zero.

The quantity $\frac{V_{\mathrm{ap}}}{V_\mathrm{p}}$ is defined as the
ratio of anti-parallel to parallel components of the measured
calorimeter transverse momentum with respect to its direction.  It is
a measure of the azimuthal balance of the event: events with one or
more high-$P_T$ particles that do not deposit energy in the
calorimeter normally have low values of
$\frac{V_{\mathrm{ap}}}{V_\mathrm{p}}$.

The quantity  $\delta$ was defined as:
\begin{equation}
\delta = \sum_i E_i(1-\cos \theta_i), \nonumber
\end{equation}

where the sum runs over energy deposits as with $P_T^{\mathrm{CAL}}$.
Due to longitudinal momentum conservation, $\delta$ peaks at twice the
electron beam energy for fully contained events.  Values of $\delta$
much larger than $2E_{e}=55\ \gev$ are usually caused by the
superposition of a NC DIS event with additional energy deposits in the
rear calorimeter not related to $ep$ collisions.

\section{Event selection}
\label{sec:elec}

 The data samples used in this analysis, the beam configurations and
 integrated luminosities, $\mathcal{L}$, are given in Table
 \ref{tab:periods}. From 2003 onwards, the electron beam was
 longitudinally polarised with average polarisation of approximately
 $\pm 30\%$. The amount of data with left- and right-handed electrons
 was approximately equal.

Offline, $P^{\mathrm{CAL}}_T$ and $P_T^{\mathrm{miss}}$ were required
to be greater than 12~GeV. The value of $P_T^{\mathrm{CAL}}$
calculated excluding the inner ring of calorimeter cells around the
forward beam-pipe hole also had to be greater than 9 GeV. These cuts
were more stringent than the corresponding online trigger
thresholds. Other preselection cuts were the requirement that the
$Z$-coordinate of the tracking vertex be reconstructed within 50 cm
(30 cm) of the nominal interaction point for 1994--2000 (2003--2007)
data and that there was a track from this vertex associated with the
lepton.  Cuts on the calorimeter timing and algorithms based on the
pattern of tracks were used to reject beam-gas, cosmic-ray and
halo-muon events.  After these preselection criteria were applied,
events with isolated electrons and muons were selected separately
using the criteria listed in Table \ref{tab:cuts}. These criteria are
described below.

In the search for isolated high-energy electrons, electromagnetic
 clusters were selected as described in Section \ref{sec:presel}. The
 distance of closest approach of the track associated with the
 electromagnetic cluster was required to be less than 10~cm.  Since
 most fake electrons are misidentified hadrons close to jets, the fake
 signal was further reduced by requiring that the electron track be
 separated by a distance, $D_{\mathrm{track}}$, of at least 0.5 units
 in $\{\eta,\phi \}$ space from other ``good'' tracks in the event. A
 track was labelled good if it had momentum larger than 0.2$\gev$, was
 associated with the event vertex and lay within $15^{\circ}< \theta <
 164^{\circ} $.  To maintain efficiency in the forward region, this
 track isolation cut was not used for {$\theta_e<45^\circ$}. Requiring
 that the matching electron track have transverse momentum greater
 than $5\gev$ also removed fake electrons. The isolated electrons were
 required to have $P^e_T> 10 \gev$ and lie within the region
 $15^{\circ} <\theta_e < 120^{\circ}$.  The fake signal rate from NC
 DIS was strongly suppressed by requiring $5 < \delta < 50$ GeV and
 further suppressed by requiring that $\phi_{\mathrm{acop}}>
 20^{\circ}$ for events that have a well defined $P_T^{X}$,
 i.e. larger than $1 \gev$ (otherwise no acoplanarity angle cut was
 applied). In addition, for low values of $P^{\mathrm{CAL}}_T$
 ($<25\gev$), where NC DIS events dominate, $\xi^2_e$ was required to
 be greater than $5000 \gev^2$. A $P_T^e$-dependent cut on
 $\frac{V_{\mathrm{ap}}}{V_\mathrm{p}}$ was applied.

\label{sec:mu}
In the search for isolated muons, the muon candidate was required to
 be isolated by a distance, $D_{\mathrm{jet}}$, of at least one unit
 in $\{\eta,\phi \}$ space from any jet with $E_T^{\rm{jet}} > 5\gev$
 and $-3 < \eta^{\rm{jet}} <3$.  The fake signal rate was reduced by
 requiring that $D_{\mathrm{track}}$ be at least 0.5. Events
 containing such isolated muon candidates with $P^{\mu}_T > 1\gev$
 were excluded from the isolated-electron search. Events in which more
 than one isolated muon with $P^{\mu}_T > 1\gev$ were found were
 rejected. The cut $\delta <70$\,GeV removed superpositions of NC DIS
 events with non-$ep$ energy deposits in the RCAL .  The muon was
 required to lie in the phase space $P^{\mu}_T > 10\gev$ and
 $15^{\circ}<\theta_{\mu}< 120^{\circ}$.  Cuts on
 $\phi_{\mathrm{acop}}$ and $\frac{V_{\mathrm{ap}}}{V_\mathrm{p}}$
 were applied to reduce the fake signal rate from dilepton production.
 The quantity $P_T^X$ was required to be greater than $12\gev$.

\section{Systematic uncertainties}
\label{sec:sys}

The major experimental sources of systematic uncertainty on the number
of events expected from SM processes originated from the luminosity
measurement, the calorimeter energy scale and the simulation of
processes in the extremities of phase space. Uncertainties on the
expectation for the observed rate of $W$ production arising from
lepton identification were negligible for the electron search and
$\pm5\%$ for the muon search.

The uncertainties on the luminosity measurements gave an overall
uncertainty of approximately $\pm2.9\%$ ($\pm3.4\%$) on the expected
SM event rate for $e^+p$ ($e^-p$) data.

The uncertainty on the CAL energy scale was investigated by globally
 scaling energy as measured in the electromagnetic section of the
 calorimeter (EMC) by $\pm2\%$.  The shifts in the expectations in the
 different $P_T^X$ bins were $\pm(0.5\text{--}3.5)\%$ in the electron
 search while they were negligible for the muon search. The hadronic
 energy-scale uncertainty was varied by globally scaling energy as
 measured in the hadronic section of the calorimeter by $\pm 3
 \%$. The effect on the SM prediction was $\pm(2\text{--}5)\%$ in both
 the electron and muon searches.

Alternative event samples were used to verify that the fake signal
rates were well simulated by the MC. The contribution of NC DIS to the
electron search was studied by selecting a sample of isolated
electrons in the phase space $\theta_e<120^{\circ}$, $P_T^e > 10$~GeV
and $P_T^{\mathrm{CAL}}>12$~GeV. The fraction of the sample arising
from NC DIS was enhanced by applying in addition the requirement that
$\delta>30$ GeV and that $\phi_{\mathrm{acop}}< 17^{\circ}$. A
systematic uncertainty of $\pm15\%$ on the fake signal rate from NC
was determined from the level of agreement between data and MC for
this selection. The effect of this uncertainty on the SM prediction
was $\pm(1\text{--}4)\%$ for the electron search and was negligible in
the muon search.

The contribution from CC DIS to the electron search arises mainly from
fake isolated electron candidates originating from the hadronic
system. To assess the ability of the MC to reproduce these events, a
sample of NC DIS candidates with additional electron candidates other
than the scattered DIS electron was selected. The additional electrons
were required to be isolated according to the same isolation criteria
as used in the isolated-lepton search. The phase space of the
additional electron was $\theta_e<120^{\circ}$ and $P_T^e > 10$~GeV.
From the agreement between data and MC, a systematic uncertainty of
$\pm25\%$ was determined for the fake signal rate from CC. The effect
of this uncertainty on the SM prediction was $\pm(2\text{--}8)\%$ for
the electron search and was negligible in the muon search.

Dilepton events produce a significant fake signal rate in the isolated
muon search. A dimuon enriched sample was selected in the phase space
$P_T^{\mathrm{CAL}}>12$ GeV, $P_T^\mu>10$~GeV and
$5^{\circ}<\theta_\mu< 120^{\circ}$.  The dimuon component was
enhanced by requiring that $\phi_{\mathrm{acop}}<20^{\circ}$ and
$\frac{V_{\mathrm{ap}}}{V_{\mathrm{p}}}<0.2$.  An uncertainty of
$\pm$25\% on the dilepton fake signal rate was determined from the
level of agreement between data and MC. The effect of this uncertainty
on the SM prediction was $\pm(4\text{--}6)\%$ for the muon search and
was negligible in the electron search.

The theoretical uncertainty of $\pm15\%$ on the $W$ production cross
section gave the largest uncertainty on the total SM prediction in
both searches, being approximately $\pm12\%$ in the muon search and
$\pm(8\text{--}12)\%$ for the electron search.

The total systematic uncertainty on the SM prediction was obtained by
summing all of the individual effects in quadrature. It was
$\pm(11\text{--}13)\%$ for the various $P_T^X$ bins in the electron
search and was $\pm (14\text{--}15\%)$ in the muon search.

\section{Isolated-lepton search results}
\label{sec:searchres}

\begin{figure}[p]
\begin{center}

\includegraphics*[scale=1.4]{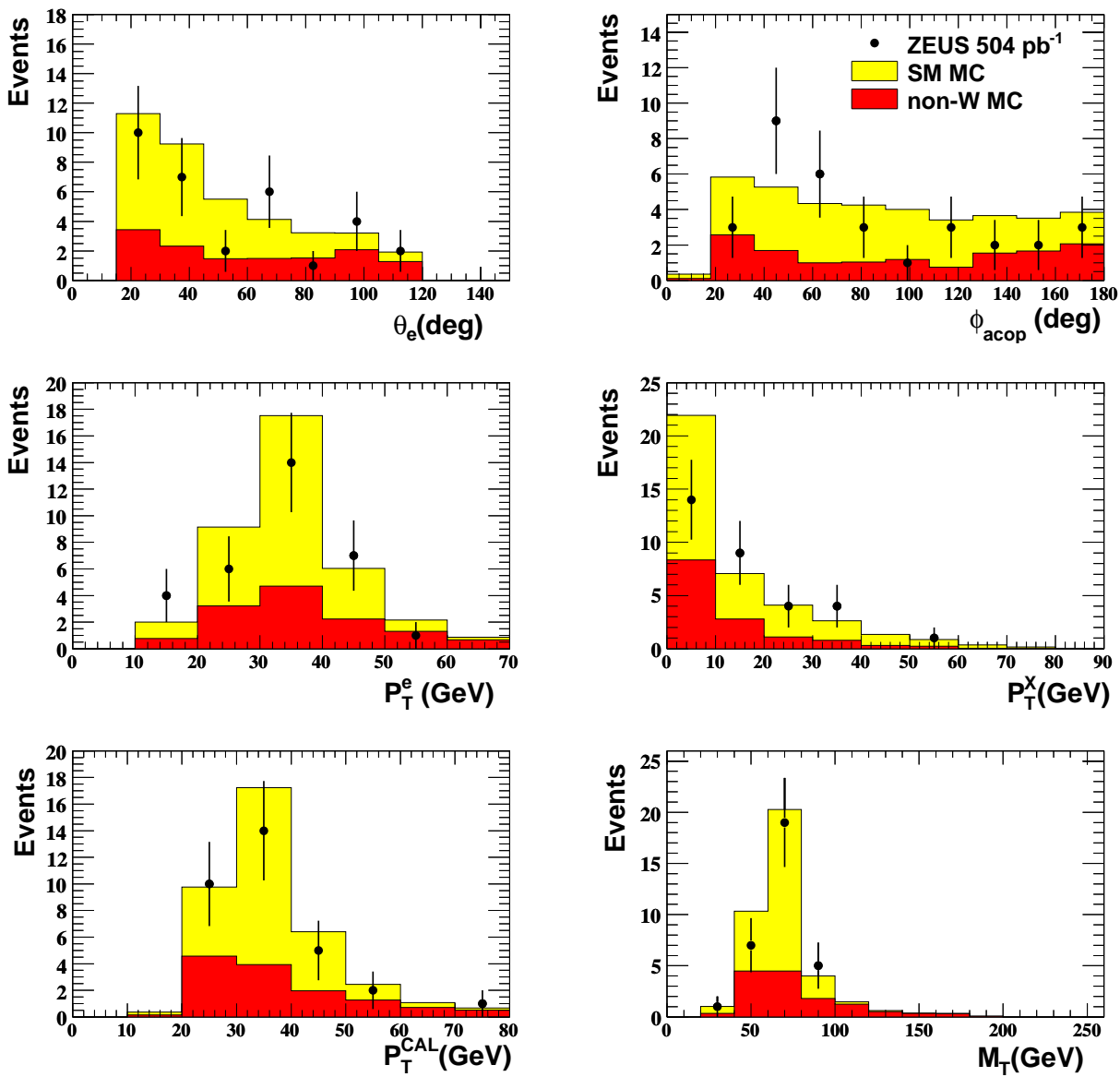}
\caption{ Distributions of the isolated electron events (points)
compared to the SM expectation
 for the $e^\pm p$ collision data. The light-shaded histogram represents the Standard Model  MC prediction, the dark shaded area being the prediction for events not arising from  $e p \rightarrow lWX $. The error bars on the data points correspond to $\sqrt{N}$ where $N$ is the number of events. The variables shown are described in detail in  the text.
\label{fig:etotcontrol}}
\end{center}
\end{figure}

Distributions of $\theta_e$, $P_T^e$, $M_T$, $\phi_{\mathrm{acop}}$,
$P_T^X$ and $P_T^{\mathrm{CAL}}$ for the isolated electron sample are
compared to the expectation from the MC simulation normalised to the
luminosity in Fig. \ref{fig:etotcontrol}.  The data are well described
by the SM Monte Carlo predictions. This is also the case when the data
are separated into $e^+p$ collision and $e^-p$ collision samples.  The
expectation from the SM and the fraction arising from $W$ production,
in bins of $P_T^X$ for the electron search are given in Table
\ref{tab:elnumbers98990405}.  No significant excess over the SM
predictions is observed.

\begin{figure}
\begin{center}
\includegraphics*[scale=1.4]{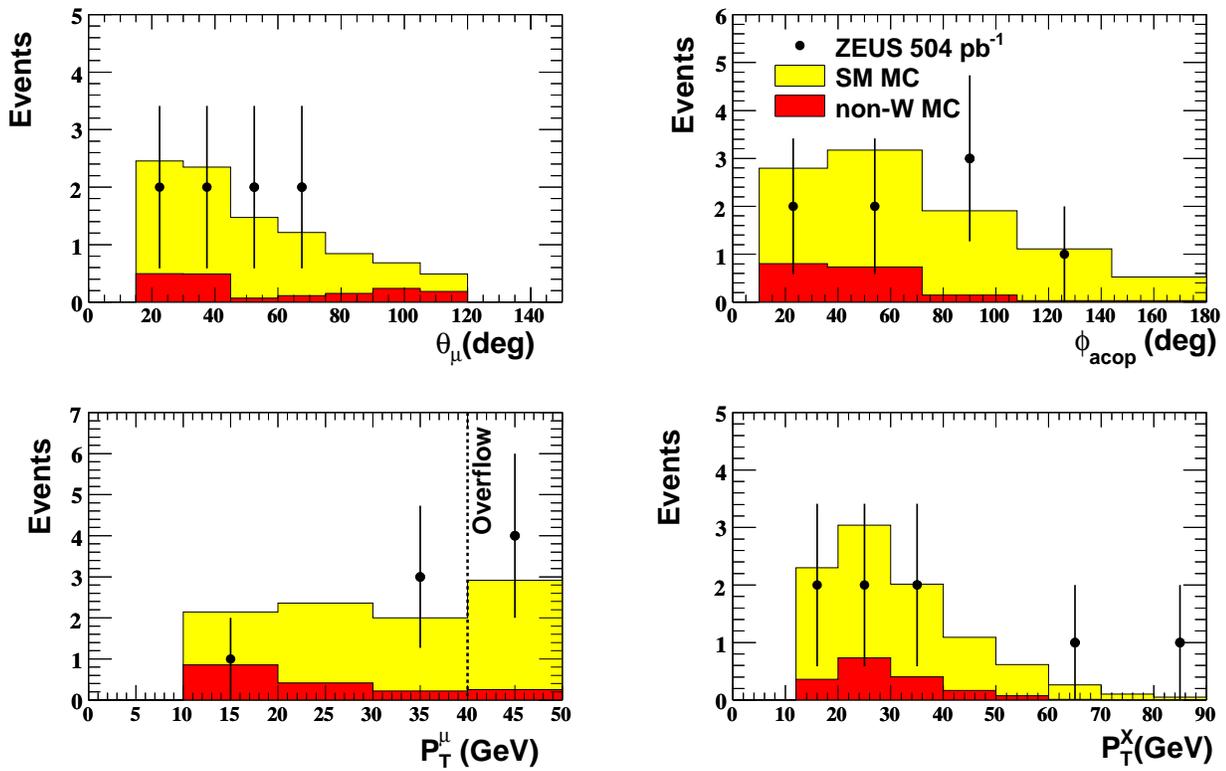}
\caption{Isolated muon data (points) compared to the SM  expectation for the $e^\pm p$ collision data. For $P_T^{\mu}> 50\,\mathrm{GeV}$ the resolution of tracking used becomes significant compared to the bin width, such events have been grouped together in the overflow bin in this figure. Other details as in the caption to  Fig. \ref{fig:etotcontrol}.\label{fig:mucontrol.pre.ep}}
\end{center}
\end{figure}

Distributions of $\theta_{\mu}$, $P_T^{\mu}$, $\phi_{\mathrm{acop}}$
and $P_T^X$ for the isolated muon sample are compared to the
expectation from the MC simulation normalised to the luminosity in
Fig. \ref{fig:mucontrol.pre.ep}.  The data are well described by the
SM Monte Carlo predictions. This is also the case when the data are
separated into $e^+p$ collision and $e^-p$ collision samples.  The
expectation from the SM and the fraction arising from $W$ production,
in bins of $P_T^X$, are given in Table \ref{tab:munumbers98-05}.

The muon and electron search results are combined in Table
\ref{tab:lepnumbers98-05}.  No excess over the SM predictions is
observed. The good agreement between the SM predictions and observed
data makes it possible to extract the $W$ production cross section.

\section{Extraction of $\mathbf{W}$ production cross section}

In order to enhance the fraction of events from $W$ production in the
electron search, an additional requirement of $\theta_e<90^{\circ}$
was applied to the sample of Section~\ref{sec:elec}. For $e^-p$
($e^+p$) collisions, this cut removed $3$ ($3$) events from data
compared to an SM expectation of $2.6$ ($2.5$).  This final sample and
the $\mu$ sample from Section \ref{sec:mu} were used to measure the
cross section for the process $ep \rightarrow lWX$ assuming a
branching fraction, $\mathcal{BR}$ ($W\rightarrow l \nu_l$), of 
10.8\,\%\cite{tpdg} per lepton.  The {\sc{Epvec}} MC reweighted as
described in Section \ref{sec:SMexp} was used in the unfolding process
to calculate acceptances.

 In the $W \rightarrow e \nu_e$ channel, the measured phase space is
$15^{\circ} < \theta_e < 90^{\circ} $, $P_T^e >10 \ \mathrm{GeV}$ and
$P_T^{\mathrm{miss}}>12\ \mathrm{GeV}$. In the $W \rightarrow \mu
\nu_\mu$ channel, the measured phase space is $15^{\circ} <
\theta_{\mu} < 120^{\circ} $, $P_T^{\mu} >10 \ \mathrm{GeV}$,
$P_T^{\mathrm{miss}}>12\ \mathrm{GeV}$ and $P_T^X >12 \ \mathrm{GeV}$.
The efficiency in the measured phase space in the $W \rightarrow e
\nu_e$ ($W \rightarrow \mu \nu_\mu$) channel is $55\%$ ($40\%$). The
total acceptance, $A_i$, for each channel is given by an extrapolation
factor from the measured phase space to the full $ep \rightarrow lWX$
phase space multiplied by the efficiency for reconstructing an event
within the measured phase space.  The acceptance for the $W
\rightarrow e\nu_e$ ($W \rightarrow \mu \nu_{\mu}$) channel was $33\%$
($11\%$). The small contribution from $W\rightarrow \tau\nu_{\tau}$
decays was taken into account.

The cross section was determined from the likelihood for observing
$n_i$ events in each search channel, defined by:
\begin{equation}
L(\sigma)=  \prod\limits_i \left(\alpha_i \int\limits_0^{\infty}{dx G_i(x)} \frac{e^{(-x)}x^{n_i}}{(n_i!)}\right) \nonumber,
\end{equation}
where the product runs over all samples being combined and $G_i(x)$ is
a Gaussian function centred on $m_i$ with width $\delta_i$;
$m_i=b_i(x)+A_i\mathcal{BR}_i\mathcal{L}\sigma$, where $b_i$ is the
number of events expected from the background and $\mathcal{BR}_i$ is
the branching ratio for the channel.  The quantity $\delta_i$ is the
statistical uncertainty on the background prediction and $\alpha_i
=(\int\limits_0^{\infty}{dx G_i(x)})^{-1}$.  The measured value of the
cross-section, $\sigma_{\mathrm{meas}}$, is that which minimises $-\ln
L(\sigma)$.  The upper and lower bounds on $\sigma_{\mathrm{meas}}$
are the values at which \mbox{$-\ln L(\sigma)=-\ln
L(\sigma_{\mathrm{meas}})+0.5$}.  Cross sections for the exclusive $W
\rightarrow e \nu_e$ and $W \rightarrow \mu \nu_{\mu}$ decay channels
were evaluated by combining $e^+p$ and $e^-p$ samples in the same
manner.

Systematic uncertainties on the extracted cross section were evaluated
by considering the effects discussed in Section \ref{sec:sys}.  In
addition, the extrapolation factor for the muon channel is sensitive
to the $P_T^X$ distribution. In order to take this into account, the
cross section in the $P_T^X<12\ \mathrm{GeV}$ region was varied by the
theoretical uncertainty on the total cross section, $\pm 15\%$,
leading to variations in the extrapolation factor of $\pm 9\%$.  The
variation observed on the combined cross section from this change was
$\pm3\%$. The systematic uncertainties from individual effects were
added in quadrature.  The dominant contribution to the total
systematic uncertainty came from the uncertainty on the fake signal
rate from CC DIS; this contributed uncertainties of about $\pm11\%$
($\pm5\%$) to the cross section for $e^-p$ ($e^+p$) collisions.

The cross sections are given in Table~\ref{tab:sigmas}.  The cross
section is given at the luminosity-weighted mean of $\sqrt{s}$ for the
data samples used.  The mean polarisation of the electron beam in the
$e^-p$ and $e^+p$ data sets is less than $3\%$. The effect of such
levels of polarisation on the inclusive cross section, $\sigma_{ep
\rightarrow lWX}$, is predicted by {\sc Epvec} to be less than $1\%$
and was neglected. The cross section is therefore quoted for a mean
polarisation of 0. When $e^+p$ and $e^-p$ data are combined the cross section is quoted for the luminosity-weighted mean of the $e^+p$ and $e^-p$ cross sections. The measured cross sections are consistent with
the SM predictions.  The statistical significance of the $\sigma_{ep
\rightarrow lWX}$ measurement was evaluated by considering the
probability of measuring an equal or larger cross section in data for
a prediction containing no $W$ production. When the sytematic
uncertainties were (were not) taken into account this probability was
$1.1\times 10^{-6}$ ($1.1\times 10^{-7}$), corresponding to a
significance of $4.7 \sigma$ ($5.2 \sigma$). The full likelihood curve
including both systematic and statistical uncertainties is available
in Appendix \ref{app:fld} and from the ZEUS web page\cite{weblike}.

\section{Summary}
A search was made for isolated high-energy electrons and muons in events with large $P_T^{\mathrm{miss}}$, compatible with single $W$ production with subsequent decay $W \rightarrow e\nu_e$ or $W \rightarrow \mu \nu_{\mu}$
in $ep$ collisions at a centre-of-mass energy of about $320 \gev$. A data sample with a total integrated 
luminosity of $504$ $\mathrm{pb}^{-1}$ was used. The rate 
of production of such events at high hadronic transverse momentum was found to
be consistent with the SM predictions. The excess in
these types of events observed by the H1 collaboration is not confirmed.
The total cross section for single $W$ production was measured to be 
\begin{equation}
\sigma_{ep\rightarrow lWX}  = 0.89^{+0.25}_{-0.22}\,{\mathrm{(stat.)}}\,\pm{0.10} {\mathrm{(syst.)}}\, \mathrm{pb} \nonumber,
\end{equation}
consistent with SM predictions. The measurement represents a four-fold improvement in precision relative to the previously published ZEUS value. This measurement constitutes strong evidence for  $W$  production in $ep$ collisions at HERA with a significance of $4.7\sigma$. 

\section*{Acknowledgements}
We are grateful to the DESY directorate for their strong support and 
encouragement. We thank the HERA machine group whose outstanding efforts 
were essential for the successful completion of this work. The design,
construction and installation of the ZEUS detector were made possible by
the efforts of many people not listed as authors.

\providecommand{\etal}{et al.\xspace}
\providecommand{\coll}{Coll.\xspace}
\catcode`\@=11
\def\@bibitem#1{%
\ifmc@bstsupport
  \mc@iftail{#1}%
    {;\newline\ignorespaces}%
    {\ifmc@first\else.\fi\orig@bibitem{#1}}
  \mc@firstfalse
\else
  \mc@iftail{#1}%
    {\ignorespaces}%
    {\orig@bibitem{#1}}%
\fi}%
\catcode`\@=12

\newpage
\pagebreak[4]

\begin{table}
\begin{center}
\begin{tabular}{|c|c|c|c|c|c|}
\hline
 &  & Electron energy & Proton energy & $\sqrt{s}$ & $\mathcal{L}$\\
 \raisebox{1.5ex}[0pt]{Period} & \raisebox{1.5ex}[0pt]{Beams}  &  (GeV) & (GeV) &  (GeV) & ($\mathrm{pb}^{-1}$) \\  
\hline
1994--1997 & $e^+p$ & $27.5$ & $820$ & $300$ & $48.2$ \\
1998--1999 & $e^-p$ & $27.5$ & $920$ & $318$ & $16.7$ \\
1999--2000 & $e^+p$ & $27.5$ & $920$ & $318$ & $65.1$ \\
\hline
2003--2004 & $e^+p$ & $27.5$ & $920$ & $318$ & $40.8$ \\
2004--2006 & $e^-p$ & $27.5$ & $920$ & $318$ & $190.9$ \\
2006--2007 & $e^+p$ & $27.5$ & $920$ & $318$ & $142.4$ \\

\hline
\end{tabular}
\end{center}
\caption{Details of the different data subsamples over the 1994--2007 running period.  From  2003 onwards the electron beam was longitudinally polarised.
 \label{tab:periods}}
\end{table}

\begin{table}
\begin{center}
\begin{tabular}{|c||c|c|}
\hline
Variable & Electron & Muon \\
\hline
\hline
$P^{\mathrm{CAL}}_T$ & $>12\gev$ & $>12\gev$ \\
\hline
$P_T^{\mathrm{miss}}$ & $>12\gev$ & $>12\gev$ \\
\hline
$D_{\mathrm{track}}$ & $>0.5$ for $\theta_e>45^{\circ}$ & $>0.5$ \\
\hline
$P_T^l$ &$>10\gev$ & $>10\gev$ \\
\hline
$\theta_l$ & $15^{\circ}<\theta_e < 120^{\circ}$ &$15^{\circ}<\theta_\mu < 120^{\circ}$  \\
\hline
$\delta$ & $5< \delta <50\gev$& $<70\gev$\\
\hline
$\phi_{\mathrm{acop}}$ &  $>20^{\circ}$& $>10^{\circ}$ \\
\hline
 $\xi^2_l$ & $>5000\gev^2$ for $P_T<25\gev$  &  --- \\
\hline
$\frac{V_{\mathrm{ap}}}{V_{p}}$ & $<0.5$ ($<0.15$ for $P_T^e<25\gev$) &  $<0.5$ ($<0.15$ for $P^{\mathrm{CAL}}_T<25\gev$) \\
\hline
$D_{\mathrm{jet}}$ & implicit &  $>1.0$\\
\hline
\# isolated $\mu$ &  0 & 1 \\
\hline
$P_T^X$ & --- & $>12\gev$ \\
\hline
\end{tabular}
\end{center}
\caption{Selection criteria for the isolated electron and muon searches. \label{tab:cuts}}

\end{table}

\begin{center}
\begin{table}

\begin{tabular}{|c|c|c|c|}
 \hline
   Isolated $e$  & &   &   \\  
 Candidates & \raisebox{1.5ex}[0pt]{$P_T^X < 12$ GeV}  & \raisebox{1.5ex}[0pt]{$ 12 < P_T^X < 25$ GeV} &  \raisebox{1.5ex}[0pt]{$P_T^X > 25$ GeV} \\  
 \hline
 \hline
$e^{-}p \ 208 \ \mathrm{pb^{-1}} $ & $9/11.3 \pm 1.5 \ (54\%) $ & $5/3.4 \pm 0.5 \ (62\%) $ & $ 3/3.2 \pm 0.5 \ (69\%) $ \\
$e^{+}p \ 296 \ \mathrm{pb^{-1}} $ & $7/12.6 \pm 1.7 \ (68\%) $ & $5/3.9 \pm 0.6 \ (72\%) $ & $ 3/4.0 \pm 0.6 \ (77\%) $ \\
 \hline
 \hline
$e^{\pm}p \ 504 \ \mathrm{pb^{-1}} $ & $16/23.9 \pm 3.1 \ (61\%) $ & $10/7.4 \pm 1.0 \ (67\%) $ & $ 6/7.3 \pm 1.0 \ (73\%) $ \\
 \hline
\end{tabular}

\caption{Results of the search for events with isolated 
electrons and missing transverse momentum. The number of observed events
is compared to the SM prediction (observed/expected). The fraction of the SM expectation arising from W production is given as a
percentage in parentheses. The quoted errors contain statistical and
systematic uncertainties added in quadrature.
\label{tab:elnumbers98990405}}
\end{table}
\end{center}

\begin{table}[p]

\begin{center} 
\begin{tabular}{|c|c|c|}
 \hline
 Isolated $\mu$  &  &   \\  
 Candidates &  \raisebox{1.5ex}[0pt]{$ 12 < P_T^X < 25$ GeV} &\raisebox{1.5ex}[0pt]{$P_T^X > 25$ GeV} \\  
 \hline
 \hline
$e^{-}p \ 208 \ \mathrm{pb^{-1}} $ & $1/1.6 \pm 0.3 \ (77\%) $ & $ 2/2.3 \pm 0.4 \ (85\%) $ \\
$e^{+}p \ 296 \ \mathrm{pb^{-1}} $ & $2/2.2 \pm 0.3 \ (82\%) $ & $ 3/3.4 \pm 0.5 \ (81\%) $ \\
 \hline
 \hline
$e^{\pm}p \ 504 \ \mathrm{pb^{-1}} $ & $3/3.8 \pm 0.6 \ (80\%) $ & $ 5/5.7 \pm 0.8 \ (83\%) $ \\
 \hline
\end{tabular}
\end{center}
\caption{Results of the search for events with isolated muons and
missing transverse momentum. Other details as in the caption to Table \ref{tab:elnumbers98990405}. \label{tab:munumbers98-05}}
\end{table}

\begin{table}[p]
\begin{center} 

\begin{tabular}{|c|c|c|c|}
 \hline
 Isolated Lepton  & &  &   \\  
 Candidates & \raisebox{1.5ex}[0pt]{$P_T^X < 12$ GeV} & \raisebox{1.5ex}[0pt]{$ 12 < P_T^X < 25$ GeV} & \raisebox{1.5ex}[0pt]{ $P_T^X > 25$ GeV} \\  
 \hline
 \hline
$e^{-}p \ 208 \ \mathrm{pb^{-1}} $ & $9/11.3 \pm 1.5 \ (54\%) $ & $6/5.1 \pm 0.7 \ (67\%) $ & $ 5/5.5 \pm 0.8 \ (75\%) $ \\
$e^{+}p \ 296 \ \mathrm{pb^{-1}} $ & $7/12.6 \pm 1.7 \ (68\%) $ & $7/6.2 \pm 0.9 \ (75\%) $ & $ 6/7.4 \pm 1.0 \ (79\%) $ \\
 \hline
 \hline
$e^{\pm}p \ 504 \ \mathrm{pb^{-1}} $ & $16/23.9 \pm 3.1 \ (61\%) $ & $13/11.2 \pm 1.5 \ (71\%) $ & $ 11/12.9 \pm 1.7 \ (77\%) $ \\
 \hline
\end{tabular}

\end{center}
\caption{Results of the search for events with isolated electrons or muons and missing transverse momentum. Other details as in the caption to Table \ref{tab:elnumbers98990405}. \label{tab:lepnumbers98-05}}
\end{table}
\begin{table}
\begin{center}
\begin{tabular}{|c|c|c|c|c|}
\hline
 Process & $P_T^X>$ (GeV) & $\langle \sqrt{s} \rangle $ (GeV) &$\sigma$ (pb) & $\sigma_{\mathrm{SM}}$ (pb) \\
\hline
\hline
$ep \rightarrow lWX $ &  &  &   &  \\
$W\rightarrow e \nu_e $ & \raisebox{1.5ex}[0pt]{0} & \raisebox{1.5ex}[0pt]{316} &  \raisebox{1.5ex}[0pt]{$0.090^{+0.032}_{-0.028}\ (\mathrm{stat.})\ ^{+0.013}_{-0.013}\ (\mathrm{syst.}) $}  & \raisebox{1.5ex}[0pt]{0.13} \\
\hline
$ep \rightarrow lWX$  & &  &     &  \\
$W\rightarrow \mu \nu_{\mu} $  & \raisebox{1.5ex}[0pt]{12} & \raisebox{1.5ex}[0pt]{316} &  \raisebox{1.5ex}[0pt]{ $0.044^{+0.022}_{-0.018}\ (\mathrm{stat.})\ ^{+0.006}_{-0.006}\ (\mathrm{syst.}) $}  & \raisebox{1.5ex}[0pt]{0.05} \\
\hline
\hline
$e^+p \rightarrow lWX $  & 0 & 315 &   $0.82^{+0.31}_{-0.26}\ (\mathrm{stat.})\ ^{+0.08}_{-0.08}\ (\mathrm{syst.}) $  & 1.2 \\
$e^-p \rightarrow lWX $  & 0 & 318 &   $1.03^{+0.45}_{-0.38}\ (\mathrm{stat.})\ ^{+0.16}_{-0.16}\ (\mathrm{syst.}) $  & 1.3 \\
\hline
$ep \rightarrow lWX $  & 0  & 316 &   $0.89^{+0.25}_{-0.22}\ (\mathrm{stat.})\ ^{+0.10}_{-0.10}\ (\mathrm{syst.}) $  & 1.2 \\
\hline

\end{tabular}
\end{center}
\caption{Extracted $W$ production cross sections. The predicted SM cross section, $\sigma_{\mathrm{SM}}$, is given in the last column and has an estimated uncertainty of $\pm15\%$.
 \label{tab:sigmas}}
\end{table}

\pagebreak

\appendix

\section{Full likelihood distribution}
\label{app:fld}
\begin{figure}[h]
\begin{center}
\includegraphics*[scale=0.8]{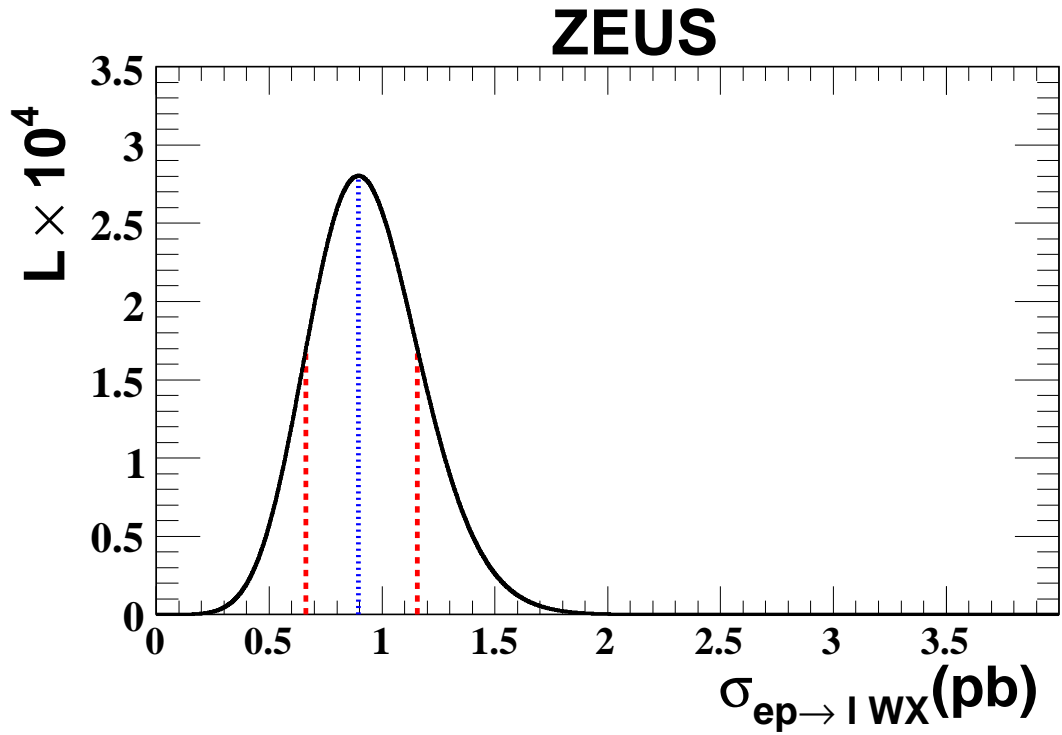}
\caption{Full likelihood distribution, including systematic uncertainties, for $\sigma_{ep \rightarrow lWX}$. The dotted line indicates the central value
of the cross section, the dashed lines indicate the upper and lower bounds on the cross section.\label{fig:fulllike}}
\end{center}
\end{figure}

\end{document}